\documentstyle[12pt,epsf]{article}

\title{\bf Radial Oscillations of Rotating Strange Stars in Strong Magnetic Fields}
\author{S.Singh\thanks{E--mail : santokh@ducos.ernet.in},
N.Chandrika Devi\thanks{E--mail : ncdevi@physics.du.ac.in},
V.K.Gupta\thanks{E--mail : vkg@ducos.ernet.in} ,
Asha Gupta and
J.D.Anand\thanks{E--mail : jda@ducos.ernet.in} \\
       {\em Department of Physics and Astrophysics,} \\
       {\em University of Delhi, Delhi-110007, India,and} \\
       {\em Inter University Centre for Astronomy and Astrophysics,} \\
       {\em Ganeshkhind, Pune-411007, India} \\
       }

\begin {document}
\baselineskip=2\baselineskip
\maketitle

\begin{abstract}
     In this paper we study radial oscillations of the rotating strange stars in strong magnetic fields in the Density Dependent Quark Mass (DDQM) model. We see that increase of frequency i.e. difference in frequency of rotating and non-rotating stars is more for higher magnetic fields. The change is small for low mass stars but it increases with the mass of the star. This change of frequency is significant for maximum mass whereas it is marginal for a 1.4$M_{\odot}$ star.

\textit{Subject headings}:magnetic fields:stars:oscillations:rotations.
\end{abstract}
\pagebreak  

\baselineskip= 15pt
\begin{section}{Introduction}
  Since the original suggestion made by Comeron [1965] more than three decades ago that vibrations of neutron stars could excite motions that can have interesting physical applications, there have been several investigations of vibrational properties of neutron stars. It has been found that for typical neutron star models with mass about one solar mass and radius about 10kms these vibrations have periods in the range (3-5)ms and this period is relatively insensitive to the exact value of central density [Cutler, Lindblom and Splenter, 1990].

   The possibility that radial oscillations of neutron stars give rise to the oscillations observed with radio subpulses of pulsars was proposed by Boriakoff [ 1976 ]. Inspite of doubts about this interpretation the possibility of detecting the radial oscillations have not been excluded [Van Horn, 1980]. More recently, periodicities have been observed in X-ray bursts [Sadeh et al, 1982] which have raised considerable interest in radial oscillations [Glass and Lindblom, 1983; Muslinov and Tsygan, 1986; Marti et al, 1988]. Recently Vath and Chanmugan [1992] have determined the two lowest frequency radial oscillation modes for several equations of state. They found that their results for neutron stars differ from those of Glass and Lindblom [1983] even if the central density is same in both the works.

    Rotation is a general property of all stellar bodies i.e. galaxies and individual stars as evidenced by Doppler broadening of their spectral lines. When the core of a massive star collapses to produce neutron star which ultimately may convert to a strange star [Witten, 1984; Farhi and Jaffe, 1984; Haensel, Zdunik and Schaeffer, 1986; Alcock, Farhi and Olinto, 1986 and Madsen and Haensel 1991] ; the conservation of angular momentum assures its enhanced rotation. Rotation has effects on structure and stability of relativistic stars and on space-time. The rotation of local inertial frames affects the internal structure of rotating stars sometimes referred as dragging of the inertial frames or as the Lense-Thirring effect. The frame dragging frequency is largest at the centre of a rotating star and falls to zero at large distances where the space-time is flat.

    As rotation and magnetic field are intrinsic properties of all stellar objects, it is intresting to ask what will be the normal modes of a rotating magnetised star. Rotation can lead to coupling between oscillation modes. This can influence the dissipation of radial vibrational energy of the star and also its rotational stability. The determination of normal mode frequencies of a rotating neutron star is non-trivial. The Newtonian formulation of this aspect of neutron stars has been studied only recently by Ipser and Lindblom [1989, 1990]. Post-Newtonian frequencies for pulsations of rapidly rotating neutron stars have been reported by Cutler and Lindblom [1992]. The radial modes of rotating neutron stars in the Chandrasekhar-Friedman [1972] formalism have been studied by Datta et al [1998].

     Since we are interested in studying the radial oscillation of magnetised rotating quark stars, we would like to mention that not much work has been done even for non-rotating  magnetised quark stars. Recently Vath and Chanmugan [1992], Gondek and Zdunik [1999], Benvenuto and Lugones [1998], Anand et al [2000a] have studied the radial oscillations of strange stars. Usually the strange stars are studied in the framework of MIT Bag Model. An alternative description of quark matter in which the confinement is treated by assuming a baryon density dependence of the quark masses was introduced by Fowler, Raha and Weiner [1981] and Plumer, Raha and Weiner [1984]. Later on, the model was reformulated [Benvenuto and Lugones; 1995a, 1995b and 1998] to show that the properties of SQM in this model are quite similar to those predicted by MIT Bag Model. Anand et al (2000b) used the Density Dependent Quark Mass (DDQM) model to study the bulk viscosity of SQM. DDQM model effectively includes the first order QCD coupling correction and yet is much easier to work with. However in some respect DDQM model is quite different from the MIT Bag model. In the low pressure region, the speed of sound is much larger than the relativistic $c/\sqrt{3}$ of the Bag model, and it agrees remarkably well with phenomenologically expected values from hadronic collisions over a large range of temperatures [Fowler et al(1981)]. Motivated by these arguments Anand et al [2000a] have studied the radial oscillation of magnetised quark stars in Density Dependent Quark Mass (DDQM) model.

    In  this paper, we study the radial oscillations of magnetised rotating strange stars in DDQM model and compare the results with Anand et al (2000a) for non- rotating stars. The plan of the paper is as follows : in section 2, a brief review of the DDQM model is given.Section 3 deals with formalism and finally in section 4, the results and discussions are presented.
\end{section}

\vspace{0.5cm}

\begin{section}{Density Dependent Quark Model in Magnetic Field}
   Following Benvenuto et al (1998) and Anand et al (2000a \& b), we shall treat SQM as a free gas of u,d,s quarks and electrons where the mass of each quark is parametrised as
\begin{equation}
   m_{u}=m_{d}=C/3n_{B}, m_{s}=m_{so}+C/3n_{B}     
\end{equation}
  Here $m_{so}$, the strange quark current mass and C are arbitrary constants, restrained by the requirement that at T=0

$(a)E/n_{B}<930Mev$ for SQM 

and                        

\begin{equation}
 (b)E/n_{B}>940Mev
\end{equation}
 for two flavour quark matter. We have selected $C=75Mev fm^{-3}$ and $m_{so}=140Mev$ well in the stability window calculated by Benvenuto and Lugones (1998).

   We consider u, d, s, e system in the presence of a magnetic field B which is directed along the Z-axis. The energy of a spin 1/2 particle of mass $m_{i}$ and charge $q_{i}$ in the presence of magnetic field is well known and is given by
\begin{equation} 
   \varepsilon_{i}=\sqrt{m_{i}^{2}+p_{z,i}^{2}+2q_{i}B(n+s+1/2)} 
\end{equation}    
where $s=+1/2 (-1/2)$ refer to spin up (down) states of the particle, and $p_{z,i}$ is the momentum along the Z-axis.We have used $h/2\pi=c=1$.
  The thermodynamical potential of the system is given by 
\begin{equation}
  \Omega=\sum_{i}\Omega_{i}=-\sum_{i}\frac{g_{i}q_{i}B}{2\pi^{2}}\sum_{n}(2-\delta_{no})\int{dp_{z}[ln{(1+e^{-\beta(\epsilon_{i}-\mu_{i})})}]}-\frac{8}{45}\pi^{2}T^{4}   
\end{equation}
  where $i=(u, d, s, e)$, $g_{i}=6$ for $(u, d, s)$ and 2 for electrons. The last term in eqn. (4) is the contribution of gluons. The expressions for pressure $p_{i}$, energy density $E_{i}$ and number density $n_{i}$ of various species are obtained by using the well known relations:
\begin{equation}
    p_{i}={(n_{B}\frac{\partial\Omega_{i}}{\partial{n_{B}}})}_{T,\mu_{i}}-\Omega_{i}
\end{equation}                                                                
\begin{equation}        
  E_{i}=-p_{i}+\mu_{i}n_{i}-T{(\frac{\partial\Omega_{i}}{\partial{T}})}_{n_{B},\mu_{i}}
\end{equation}
  and                                                            
\begin{equation}
  n_{i}={(\frac{\partial\Omega_{i}}{\partial\mu_{i}})}_{T,n_{B}}      
\end{equation} 
At T=0, the above expressions reduce to 
\begin{equation}
   p_{i}=\frac{g_{i}q_{i}B}{2\pi^{2}}\sum_{n=0}^{{n_{max}}^{i}}(2-\delta_{no})[\frac{1}{2}\mu_{i}K_{fi}-({m_{i}}^{2}+2nq_{i}B+\frac{2m_{i}C}{3n_{B}})ln(\frac{\mu_{i}+K_{fi}}{\sqrt{{m_{i}}^{2}+2nq_{i}B}})]        
\end{equation}
\begin{equation}
   E_{i}=\frac{g_{i}q_{i}B}{2\pi^{2}}\sum_{n=0}^{{n_{max}}^{i}}(2-\delta_{no})[\frac{1}{2}\mu_{i}K_{fi}+\frac{1}{2}({m_{i}}^{2}+2nq_{i}B+\frac{2m_{i}C}{3n_{B}})ln(\frac{\mu_{i}+K_{fi}}{\sqrt{{m_{i}}^{2}+2nq_{i}B}})]    
\end{equation}
 and
\begin{equation}
 n_{i}=\frac{g_{i}q_{i}B}{2\pi^{2}}\sum_{n=0}^{{n_{max}}^{i}}(2-\delta_{no})K_{fi}
\end{equation}                                                                  where 
\begin{equation}
  K_{fi}=\sqrt{{\mu_{i}}^{2}-{m_{i}}^{2}-2nq_{i}B}              
\end{equation}
 and
\begin{equation}
 {n_{max}}^{i}=int[\frac{({\mu_{i}}^{2}-{m_{i}}^{2})}{2q_{i}B}]      
\end{equation}        
 As the system is in $\beta$-equilibrium, we have
\begin{equation}
       \mu_{d}=\mu_{s}                    
\end{equation}
 and  
\begin{equation}
  \mu_{d}=\mu_{u}+\mu_{e}           
\end{equation}
 (Assuming that neutrinos/antineutrinos produced stream out freely, $\mu_{\nu}=0$). The charge neutrality condition gives
\begin{equation}
      2n_{u}-n_{d}-n_{s}-3n_{e}=0          
\end{equation}
  and the baryon number density of the system is
\begin{equation} 
      n_{B}=(n_{u}+n_{d}+n_{s})/3       
\end{equation}
 For a given baryon density and given magnetic field B the chemical potentials of various species are evaluated self-consistently by demanding charge neutrality and $\beta$-equilibrium conditions. This enables us to make a profile of pressure ($p=\sum{p_{i}}$) and energy density ($E=\sum{E_{i}}$) for various values of baryon densities and magnetic fields. (See Anand et al, 2000a).
\end{section}
\vspace{0.5cm}
\begin{section} {Radial Pulsations of a Slowly Rotating Magnetised Quark Star}
  We use the formalism given by Chandrasekhar and Friedman (1972) and Datta et al (1998) which is fully general relativistic to calculate the effect of rotation (to order $\Omega^{2}$, where $\Omega$ is the angular vlocity of rotation) on the eigenfrequencies of radial pulsations of stars. It gives an exact formula to calculate the frequency ($\sigma^{\prime}$) of oscillations of a ``slowly'' rotating stellar configuration which depends only on the knowledge of the Lagrangian displacement  associated with the radial mode of oscillation of the non-rotating configuration and the uniform spherical deformation caused by rotation. By ``slow'' rotation we mean those rates of rotation that are small in comparision to the central mass shed limit.

   The Chandrasekhar-Friedman formula is of the form
\begin{equation}
{\sigma^{\prime}}^{2}I_{1}=I_{2}+I_{3}+I_{4}          
\end{equation}

where

\begin{equation}
I_{1}=\int{d^{3}x\sqrt{-g}e^{2\lambda}\xi^{2}{u_{o}}^{2}(p+\rho{c^{2}})}
\end{equation}  

\begin{eqnarray}
I_{2} 
&=& \int{d^3}x\sqrt{-g}[\Gamma{p}{(\Delta{N}/N)}^2+\frac{1}{(\epsilon+p)}\frac{dp}{dr}\frac{d(\rho{c^{2}})}{dr}\xi^{2}  \nonumber \\
&-& 2(\frac{\Delta{N}}{N})\frac{dp}{dr}\xi -2(p+\frac{c^{4}}{8\pi{G}r^2}){(\delta\lambda)}^2]  
\end{eqnarray}

\begin{eqnarray}
I_{3} &=& \int{d^3}x\sqrt{{-g}_{s}}V^{2}[\frac{{(\Gamma{p})}^{2}}{(p+\rho{c^{2}})}{(\Delta{N}/N)}^{2}+2\Gamma{p}(\Delta{N}/N)\frac{d}{dr}(lnu_{1})\xi \nonumber \\
&+& (p+\rho{c^{2}}){(\frac{dlnu_{1}}{dr})}^{2}{\xi}^{2}+2(\delta{V}/V)\delta{p}+\frac{16\pi{G}}{c^{4}}{(p+\rho{c^{2}})}^{2}e^{2\lambda}\xi^{2} \nonumber \\
&+& 2(p+\rho{c^{2}})u_{o}u_{1}{\bar{\omega}}^{\prime}\frac{\delta\lambda}{cV^{2}}\xi]_s \end{eqnarray}

\begin{eqnarray}                              
I_{4}=\int{\frac{d^{3}xc^{2}}{16\pi{G}}r^{4}{{\sin}^{3}\theta}e^{-\lambda-\nu}{(\frac{d\bar{\omega}}{dr})}^{2}{(\delta\lambda)}^{2}}      
\end{eqnarray}

    The subscript ,s, refers to Schwarzschild (i.e. non-rotating configuration) values and $\bar{\omega}(r)$ is the angular velocity of the fluid element relative to the local inertial frame.

     In the above equations only the spherical deformation effects of rotation are manifest. Contributions due to quadrupole deformations are all taken to be zero. The metric used is similar to Hartle and Thorn (1968) i.e.
\begin{eqnarray}
    ds^{2}=-e^{2(\nu_{s}+\nu_{\Omega})}{(cdt)}^{2}+r^{2}{{\sin}^{2}\theta}{(d\phi-\omega{dt})}^{2}+r^{2}{d\theta}^{2}+e^{2(\lambda_{s}+\lambda_{\Omega})}{(dr)}^{2}                                 
\end{eqnarray}
  where $\omega(r)$ is the angular velocity of the commulative inertial frame dragging. The angular velocity of the local inertial frame is written as 
\begin{equation}
   \bar{\omega}=\Omega-\omega 
\end{equation}                                    
 where $\Omega$ is the angular velocity as seen by a distant observer. Notice that  for a non-rotating case the metric is given by
\begin{eqnarray}
  {ds}^{2}=-e^{-2\nu_{s}}{(cdt)}^{2}+e^{2\lambda_{s}}{(dr)}^{2}+{r}^{2}({d\theta}^{2}+{{\sin}^{2}\theta}{d\phi}^{2})      
\end{eqnarray}
  In eqn (22) we have written 
\begin{equation}  
   \nu=\nu_{s}+\nu_{\Omega}                                  
\end{equation} 
  and
\begin{equation}  
   \lambda={\lambda}_{s}+{\lambda}_{\Omega}                       
\end{equation}
  where ${\nu}_{\Omega}$ and ${\lambda}_{\Omega}$ are the corrections of O (${\Omega}^{2}$) due to the rotations and ${\nu}_{s}$ and ${\lambda}_{s}$ are the Scharzschild values.
 
  In Hartle's notation 
\begin{equation}
   {\nu}_{\Omega}=h_{o}                                        
\end{equation}
   and 
\begin{equation}
   {\lambda}_{\Omega}=\frac{m_{o}}{r-2Gm(r)/c^{2}}             
\end{equation}
  where $h_{o}$ and $m_{o}$ are functions of r. To  o(${\Omega}^{2}$),
\begin{equation}
     \sqrt{-g}=r^{2}e^{{(\lambda+\nu)}_{s}}[1+{(\lambda+\nu)}_{\Omega}]\sin\theta
\end{equation}
\begin{equation}   
    {\sqrt{-g}}_{s}=\frac{r^{2}e^{\nu_{s}}}{\sqrt{(1-\frac{2Gm}{rc^{2}})}}\sin\theta   
\end{equation}
\begin{equation}
   u_{o}=e^{-{\nu}_{s}}(1-{\nu}_{\Omega}+\frac{1}{2}\frac{r^{2}{\bar{\omega}}^{2}}{c^{2}}e^{-2{\nu}_{s}}{\sin}^{2}\theta)           
\end{equation}
\begin{equation}
   u_{1}=rV\sin\theta=(r^{2}\bar{\omega}/c)e^{-{\nu}_{s}}{\sin}^{2}\theta  
\end{equation}                                                                 \begin{equation}   
   V=(r\bar{\omega}/c)e^{-{\nu}_{s}}\sin\theta                       
\end{equation}

  Various other quantities appearing in equations (19)-(22) are as follows:
\begin{equation}
   \delta\lambda={\delta\lambda}_{s}+{\delta\lambda}_{\Omega}       
\end{equation}
\begin{equation}
   {\delta\lambda}_{s}=-\frac{4\pi{G}}{c^{4}}\frac{r(p+\rho{c^{2}})}{(1-\frac{2Gm}{rc^{2}})}\xi  
\end{equation}
 and
\begin{equation}
    {\delta\lambda}_{\Omega}=-\frac{4\pi{G}}{c^{4}}\frac{r(p+\rho{c^{2}})}{(1-\frac{2Gm}{rc^{2}})}[(1+\frac{p+\rho{c^{2}}}{\Gamma{p}})p_{o}+\frac{2m_{o}}{(r-\frac{2Gm}{c^{2}})}+\frac{r^{2}{\bar\omega}^{2}}{c^{2}}e^{-2{\nu}_{s}}{\sin}^{2}\theta]\xi    
\end{equation}
\begin{equation}
    \Delta{N}/N={(\Delta{N}/N)}_{s}+{(\Delta{N}/N)}_{\Omega}                
\end{equation}
  where
\begin{equation}
    {(\Delta{N}/N)}_{s}=\frac{d{\nu}_{s}}{dr}\xi-2\xi/r-\frac{d\xi}{dr}=-\frac{{\zeta}^{\prime}}{\zeta}\xi                  
\end{equation}
   where we have used $\xi=r^{-2}e^{{\nu}_{s}}\zeta$
\begin{equation}
 {(\Delta{N}/N)}_{\Omega}=[\frac{\Gamma{p}}{p+\rho{c^{2}}}{(\Delta{N}/N)}_{s}+(1/r)\xi]\frac{r^{2}{\bar{\omega}}^{2}}{c^{2}}e^{-2{\nu}_{s}}{\sin}^{2}\theta+\frac{dh_{o}}{dr}\xi                
\end{equation}
\begin{equation}
   \frac{\delta{V}}{V}=-[\frac{\Gamma{p}}{p+\rho{c^{2}}}{(\Delta{N}/N)}_{s}]-2\xi/r+\frac{d{\nu}_{s}}{dr}\xi-\frac{{\bar{\omega}}^{\prime}}{\bar{\omega}}\xi  \end{equation}
where 
\begin{equation}  
  \frac{dh_{o}}{dr}=\frac{d}{dr}[\frac{1}{3}e^{-2{\nu}_{s}}{(\frac{r\bar{\omega}}{c})}^{2}-p_{o}]
\end{equation}
\begin{equation}
   1/r^{4}\frac{d}{dr}(r^{4}j\frac{d\bar{\omega}}{dr})+4/r\frac{dj}{dr}\bar{\omega}=0
\end{equation}
\begin{equation}
   \frac{dm_{o}}{dr}=\frac{4\pi{G}{r^{2}}}{c^4}\frac{d\epsilon}{dp}(\epsilon+p)p_{o}+\frac{1}{12}\frac{j^{2}r^{4}}{c^2}{(\frac{d\bar{\omega}}{dr})}^{2}-\frac{1}{3c^2}r^{3}\frac{dj^{2}}{dr}{\bar{\omega}}^{2}
\end{equation}
\begin{eqnarray}
 \frac{dp_{o}}{dr} &=& -\frac{m_{o}(1+\frac{8\pi{G}{r^{2}}p}{c^4})}{{(r-\frac{2Gm}{c^2})}^{2}}-\frac{4\pi{G}(\epsilon+p)r^{2}}{c^4(r-\frac{2Gm}{c^2})}p_{o}+\frac{1}{12c^2}\frac{r^{4}j^{2}}{(r-\frac{2Gm}{c^2})}{(\frac{d\bar{\omega}}{dr})}^{2} \nonumber \\
&+& \frac{1}{3c^{2}}\frac{d}{dr}(\frac{r^{3}j^{2}{\bar{\omega}}^{2}}{r-\frac{2Gm}{c^2}})
\end{eqnarray}
     The pressure and energy change from
 
                  $p\rightarrow{p}+\Delta{p}$

and 

$\epsilon\rightarrow{c^{2}}\rho+c^{2}\Delta\rho$. 

Hence,
\begin{equation}
\frac{dp}{dr}\rightarrow\frac{{dp}_{s}}{dr}[1+(1+\frac{p+\rho{c^{2}}}{\Gamma{p}})p_{o}+\frac{(p+\rho{c^{2}})}{dp/dr}\frac{dp_{o}}{dr}]             
\end{equation}
\begin{equation}
    \frac{d\epsilon}{dr}\rightarrow{c^{2}}\frac{d\rho}{dr}+2(p+c^{2}\rho)\frac{\frac{(dp_{s}}{dr}+c^{2}\frac{d\rho}{dr})}{\Gamma{p}}p_{o}-\frac{{(p+\rho{c^{2}})}^{2}}{{(\Gamma{p})}^{2}}\frac{d(\Gamma{p})}{dr}p_{o}+\frac{{(p+\rho{c^{2}})}^{2}}{(\Gamma{p})}\frac{dp_{o}}{dr}               
\end{equation}

  For the sake of completeness we study the moment of inertia of a rotating strange star in strong magnetic field and the expression used is:
\begin{equation}                                                            
I=J/\Omega
\end{equation}
    where J, the angular momentum, is given by 

  $J=\frac{c^2}{6G}R^{4}(\frac{d\bar{\omega}}{dr})\vert_{(r=R)}$ 

and 

$\Omega=\bar{\omega}(R)+\frac{2GJ}{c^{2}R^{3}}$.

  The equation governing infinitesimal radial pulsations of a non-rotating star in general relativity was given by Chandrasekhar [1972] and has the following form
\begin{equation}
  F\frac{d^{2}\xi}{dr^{2}}+G\frac{d\xi}{dr}+H\xi={\sigma}^{2}\xi           
\end{equation}
 where the coefficients F, G and H are given in detail by Anand et al [2000a] and Datta et al [1998]. The equations for hydrostatic equilibrium of non-rotating degenerate stars in general relativity are
\begin{equation}
    \frac{dp}{dr}=-\frac{G(\rho+p/c^{2})(m+4\pi{r^{3}}p/c^{2})}{r^{2}(1-\frac{2Gm}{rc^{2}})}                                                               
\end{equation}   
\begin{equation}
   \frac{dm}{dr}=4\pi{r^{2}}\rho                                      
\end{equation}
\begin{equation}
     \frac{d\nu}{dr}=\frac{G(m+4\pi{r^{3}}p/c^{2})}{r^{2}c^{2}(1-\frac{2Gm}{rc^{2}})}                                                  
\end{equation}           
\begin{equation}
     \lambda=-\frac{1}{2}ln(1-\frac{2Gm}{rc^{2}})                               
\end{equation}
    The boundary conditions to solve the pulsation eqns.(48) are
 
                       $\xi(r=0)=0$   

and
\begin{equation}
    \delta{p}(r=R)=-\xi\frac{dp}{dr}-\Gamma{p}\frac{e^{\nu}}{r^{2}}\frac{\partial}{\partial{r}}(r^{2}e^{-\nu}\xi){\vert}_{r=R}=0                              \end{equation}
\end{section}
\vspace{0.5cm}         

\begin{section}{Results and Discussion}
    The calculation of the radial modes of vibration of a rotating star is straightforward as long as the rotation is supposed to be ``slow'': $\Omega<<\Omega_{c}=\sqrt{GM/R^{3}}$, but extremely tedious. For each star of the sequence corresponding to a given central density $\rho_{c}$ and magnetic field B, we have a profile of $\rho$, p, m, $\Gamma{p}$, $\nu_{s}(r)$ etc as functions of r. The solution of the eigenvalue problem has provided us $\xi(r)$, $\xi^{\prime}(r)$ for each mode.

     Our next task is to solve the second order differential equation for $\bar{\omega_{c}}$ for a given $\bar{\omega_{c}}(0)$ and obtain $\Omega$ and J after the solution is obtained. The value of $\bar{\omega_{c}}(0)$ is then adjusted to obtain each required value of $\Omega$. Next the equation for $h_{o}(r)$ and $p_{o}(r)$ are solved using the profile of $\bar{\omega}(r)$ already obtained. Once this is done all the quantities involved in the integrals $I_{1}$, $I_{2}$, $I_{3}$ and $I_{4}$ are known. The value of the frequency of radial oscillation $\sigma^{\prime}$ for the rotating star can thus be found.

   In fig.1 we plot the moment of inertia I vs $M/M_{\odot}$ for both relativistic and non-relativistic strange stars for eB=0 and $10^{5}{Mev}^{2}$. Here curves a and b correspond to non-relativistic case for eB=0 and $10^{5}{Mev}^{2}$ respectively while $a^{\prime}$, $b^{\prime}$ correspond to relativistic case for the same magnetic fields. It is seen that I increases with magnetic field for the same mass configuration for both non-relativistic and relativistic case. In fig.2 we have plotted $\omega/\Omega$ vs r/R for $M=M_{maximum}$ and for eB=0, $1\times10^{5}$, $2\times10^{5}{Mev}^{2}$ corresponding to a, b, c curves respectively. It is seen that the effect of rotation decreases as one moves from the centre to the surface of the star. However the magnetic field slows down the rotation considerably from the centre towards the surface of the star implying that the magnetised strange stars are less prone to the rotational instabilities.

    In figs.3 to 5 we show the behaviour of $\sigma$ and $\sigma^{\prime}$ (i.e. radial frequencies for non-rotating and rotating magnetised stars) for n=0, 1, 2 modes respectively for $\Omega$=0.1$\Omega_{c}$. It is interesting to note that for non-rotating strange stars the radial oscillation frequencies decrease with increasing $\rho_{c}$. For rotating stars the frequencies initially decrease but then start increasing. The increase is more pronounced for n=0 mode implying the increased stability of the magnetised strange stars.

  Figs.6 to 8 show the variation of $\sigma$ and $\sigma^{\prime}$ with $M/M_{\odot}$ and depict a behaviour similar to the one in figs.3 to 5.

   In figs.9-11, we show the plot of ${\sigma^{\prime}}^{2}-{\sigma}^{2}$ vs $\Omega/\Omega_{c}$ for eB=0, $1\times10^{5}$, $2\times10^{5}{Mev}^{2}$ for n=0, 1, 2 modes respectively. It is noted that the increase in ${\sigma^{\prime}}^{2}-{\sigma}^{2}$ is monotonic with rotation and this effect increases with increasing magnetic field.

     To conclude we find that in the presence of a strong magnetic field the rotating strange stars are less prone to rotational instabilities implying they become more stable.
\end{section}
\pagebreak

\pagebreak

\begin{figure}[ht]
\vskip 15truecm
\includegraphics{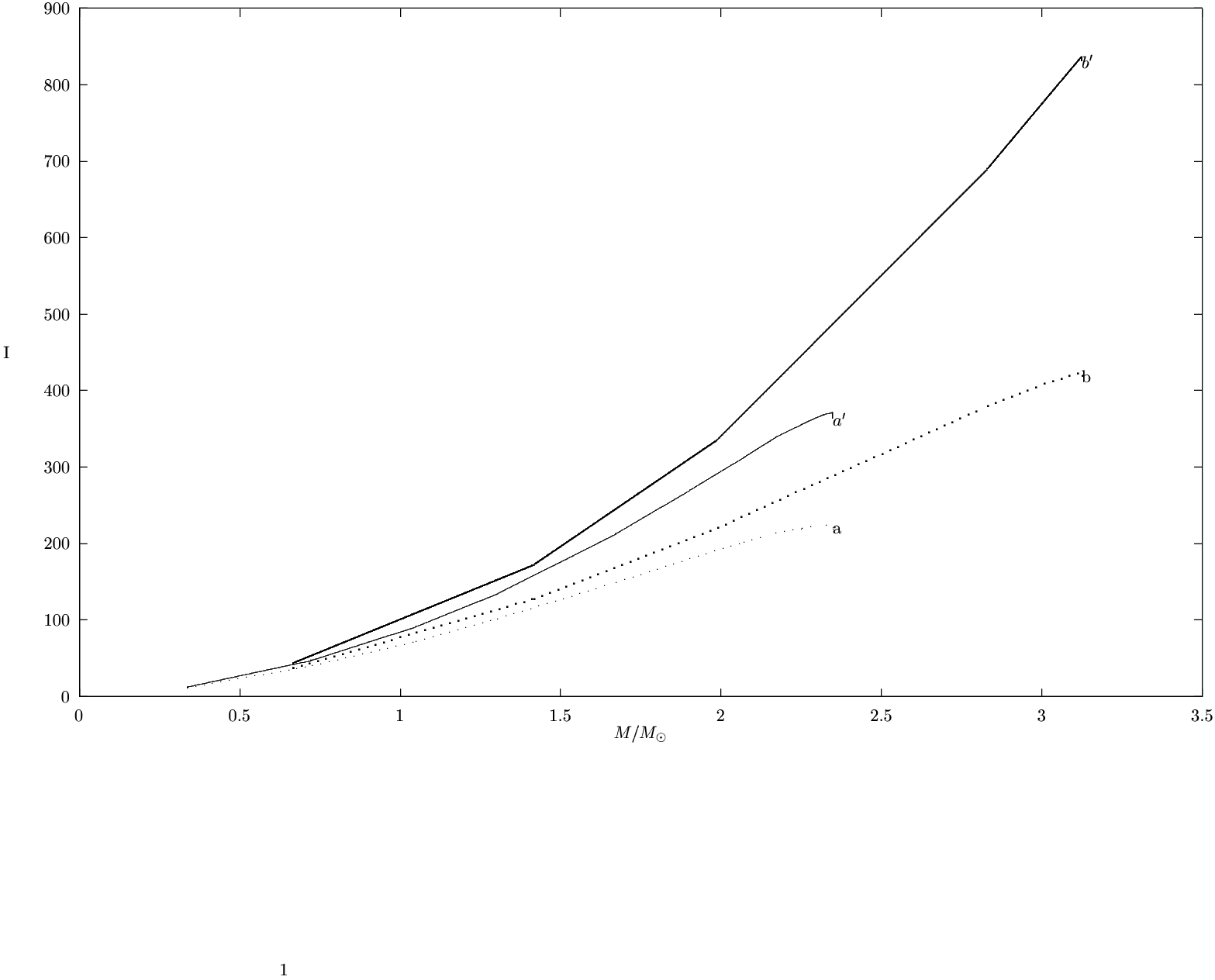}
\caption{Moment of inertia, I, in units of $M_{\odot}Km^2$ vs $M/M_{\odot}$ for both relativistic and non-relativistic strange stars for $eB=0$, $1\times10^{5}Mev^{2}$. Curves a and b correspond to the nonrelativistic case for $eB=0$ \& $1\times10^{5}Mev^{2}$ respectively while $a^{\prime}$ and $b^{\prime}$ correspond to the relativistic case.}
\end{figure}

\begin{figure}[ht]
\vskip 15truecm
\includegraphics{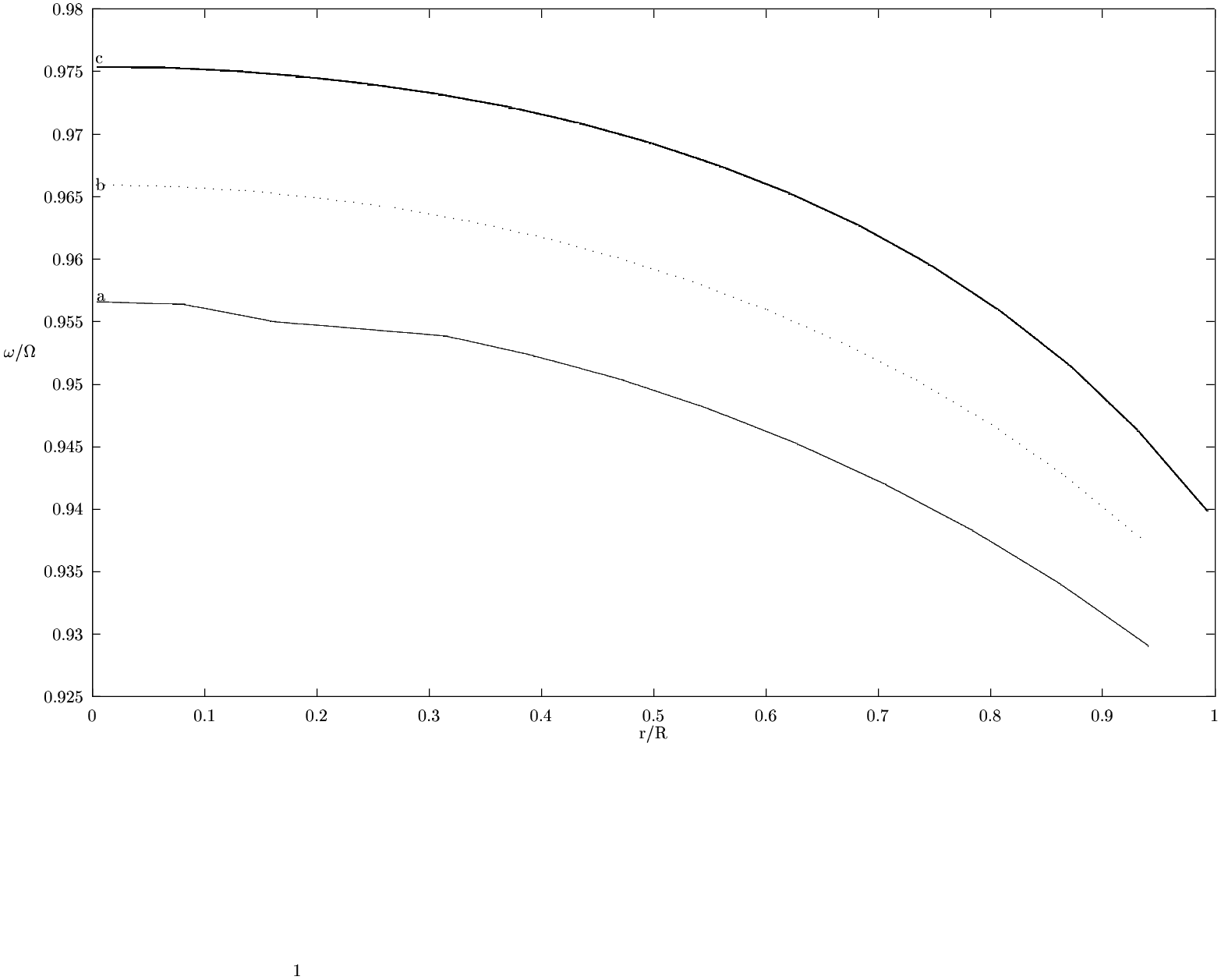}
\caption{$\omega/\Omega$ vs $r/R$ for maximum masses at $eB=0$, $1\times10^{5}$and $2\times10^{5}Mev^{2}$ corresponding to curves a,b and c respectively.}
\end{figure}

\begin{figure}[ht]
\vskip 15truecm
\includegraphics{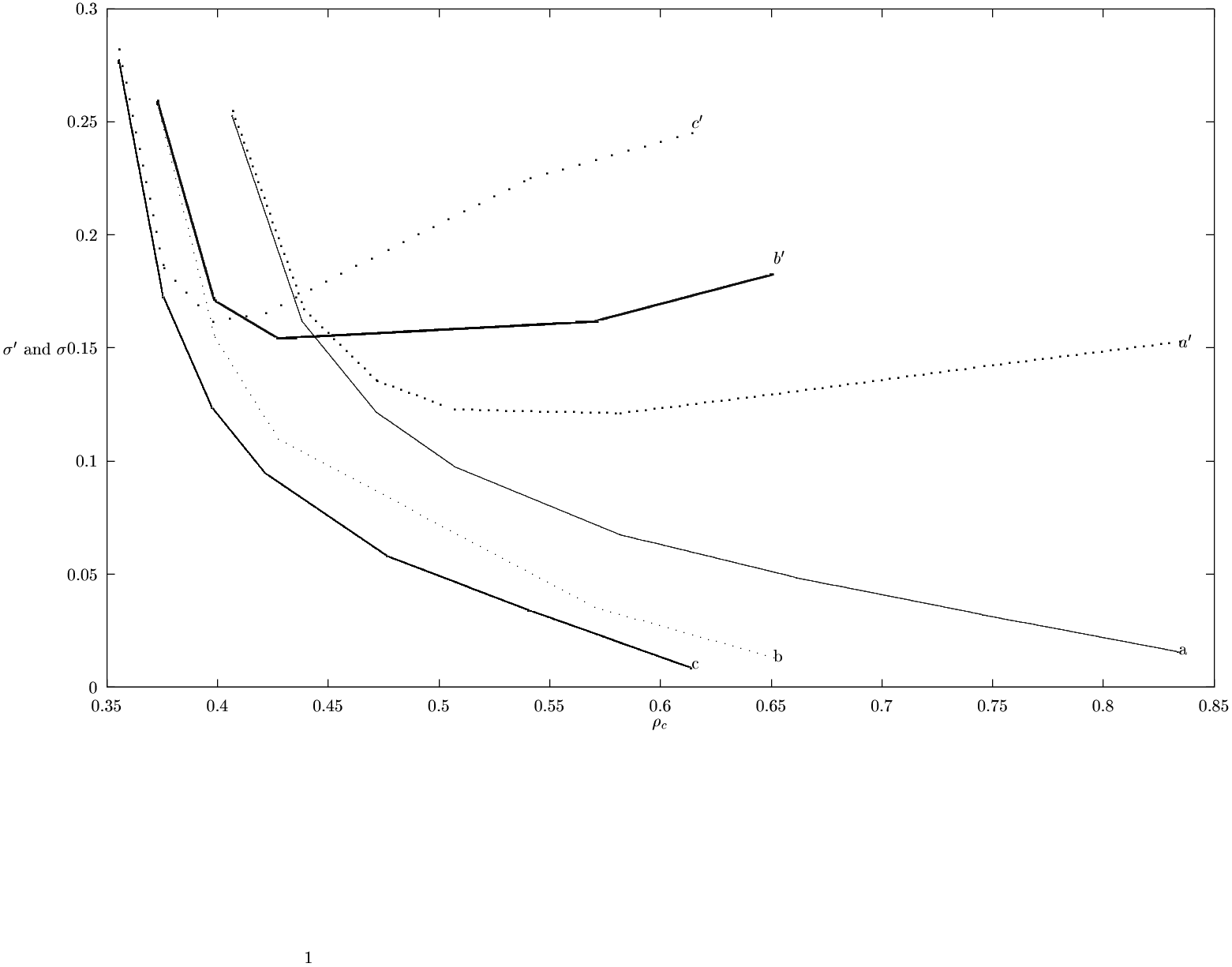}
\caption{Radial oscillation frequencies $\sigma$ (non-rotating stars) and $\sigma^{\prime}$ (rotating stars) in units of $3\times10^{5}sec.^{-1}$ vs $\rho_c$ in units of $10^{15}gm./cm.^{3}$. Curves a, b and c are for nonrotating stars while $a^{\prime}$, $b^{\prime}$ and $c^{\prime}$ refer to rotating stars for magnetic fields 0, $1\times10^{5}$ and $2\times10^{5}Mev^{2}$ respectively for $n=0$ and $\Omega=0.1\Omega_{c}$.}
\end{figure}

\begin{figure}[ht]
\vskip 15truecm
\includegraphics{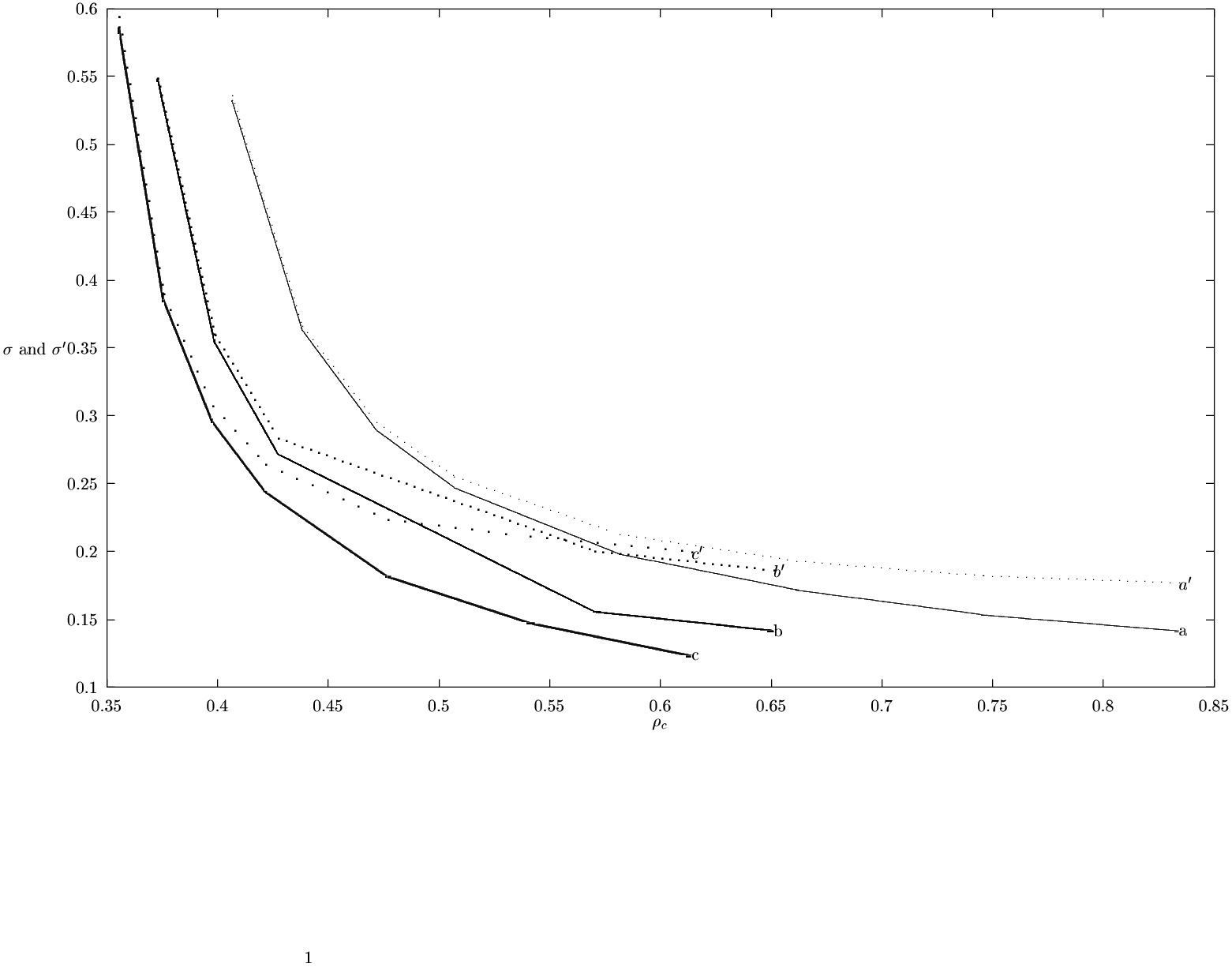}
\caption{Same as figure 3 for $n=1$.}
\end{figure}

\begin{figure}[ht]
\vskip 15truecm
\includegraphics{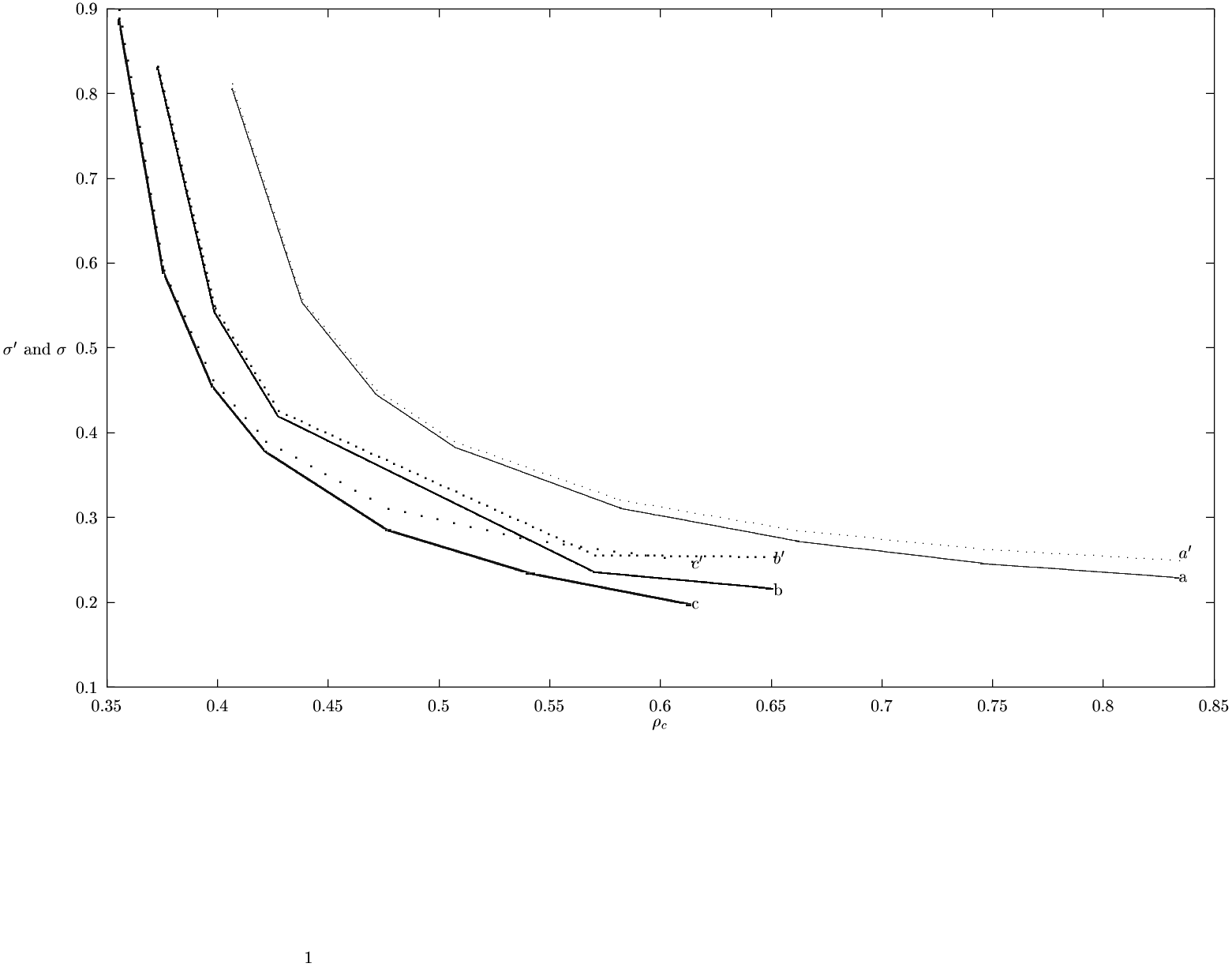}
\caption{Same as figure 3 for $n=2$.}
\end{figure}

\begin{figure}[ht]
\vskip 15truecm
\includegraphics{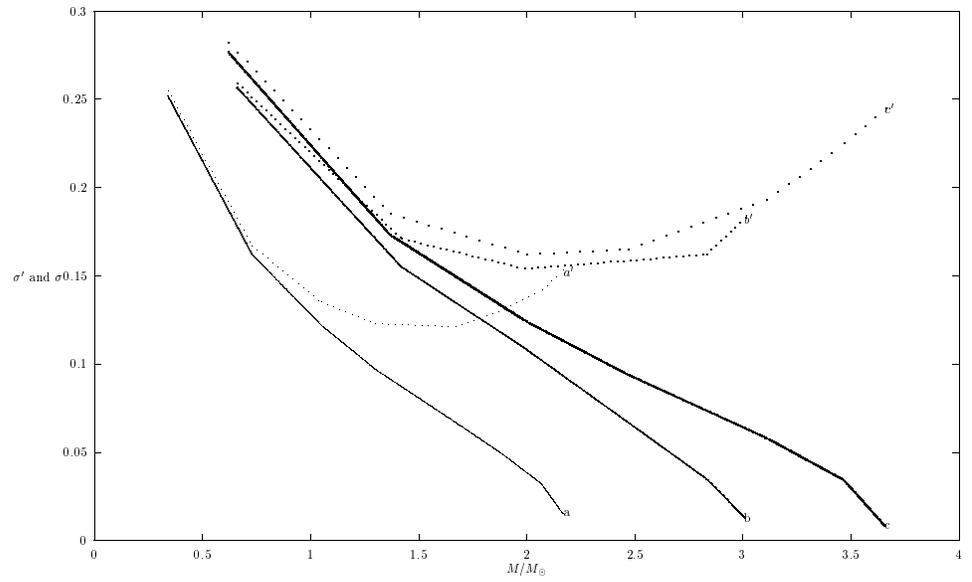}
\caption{The radial frequencies $\sigma$ and $\sigma^{\prime}$ vs $M/M_{\odot}$ for $n=0$ mode and $\Omega/\Omega_{c}$=0.1. Curves a, b, c and $a^{\prime}$, $b^{\prime}$, $c^{\prime}$ have the same meanings as in fig.3.}
\end{figure}

\begin{figure}[ht]
\vskip 15truecm
\includegraphics{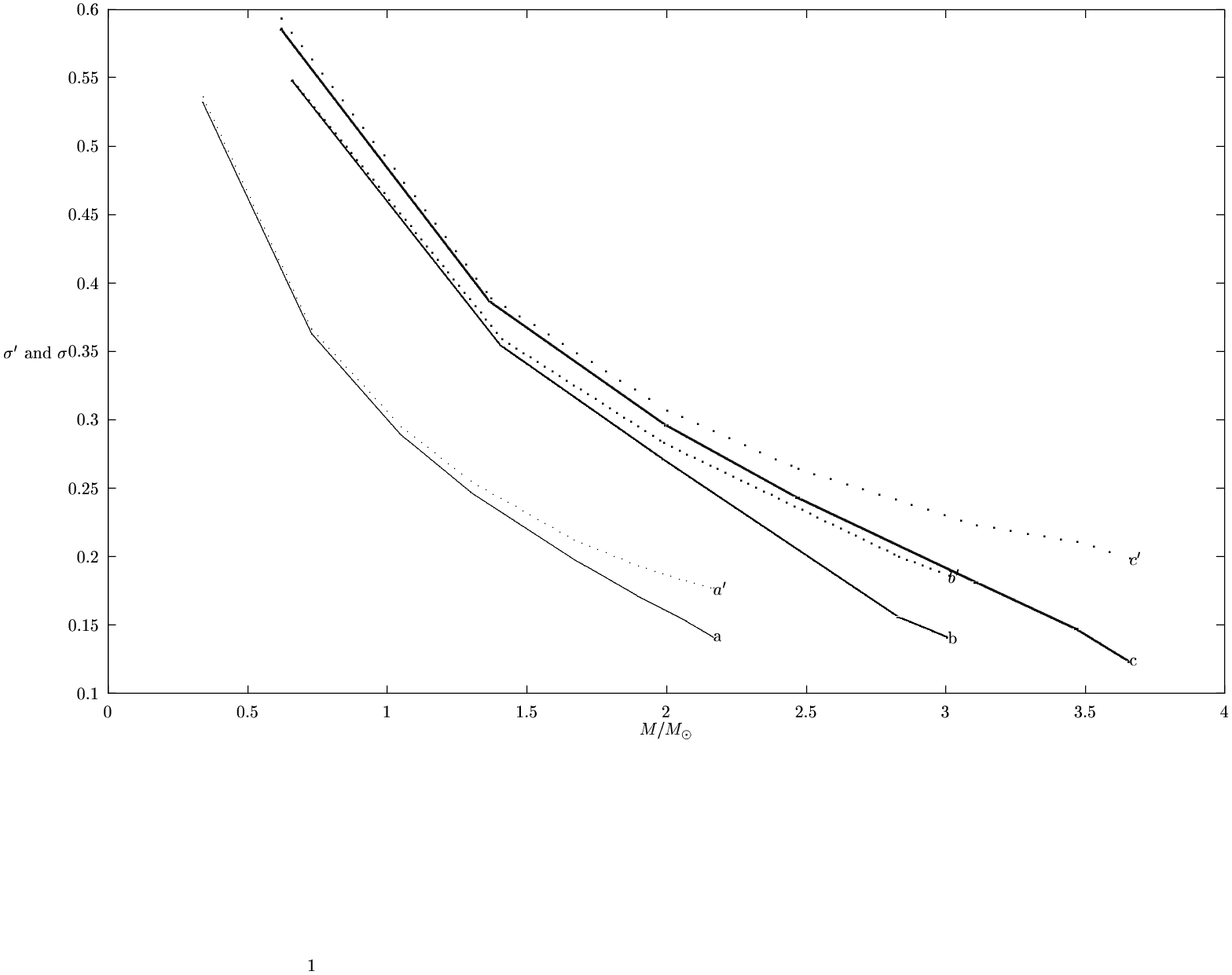}
\caption{Same as figure 6 for $n=1$.}
\end{figure}

\begin{figure}[ht]
\vskip 15truecm
\includegraphics{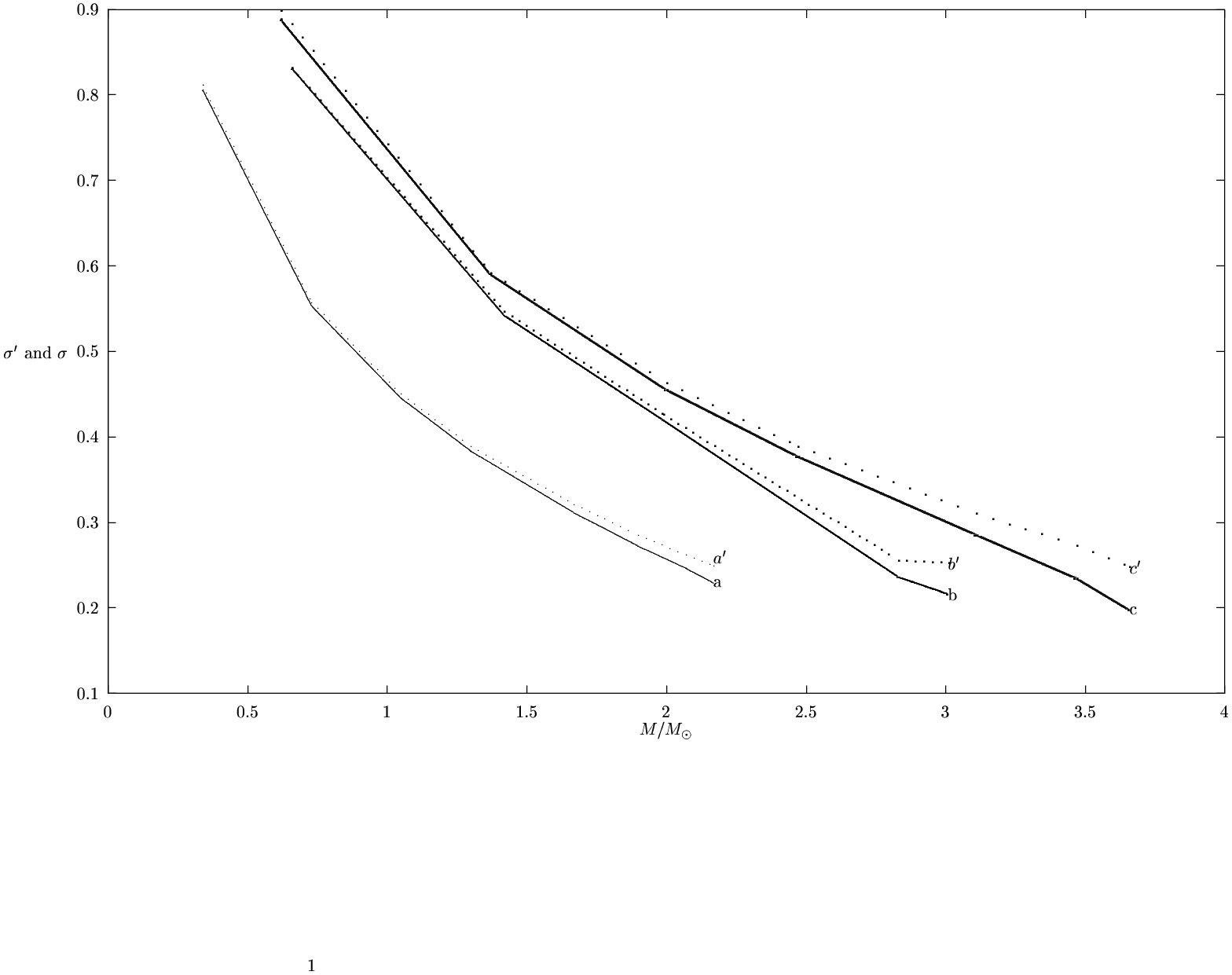}
\caption{Same as figure 6 for $n=2$.}
\end{figure}

\begin{figure}[ht]
\vskip 15truecm
\includegraphics{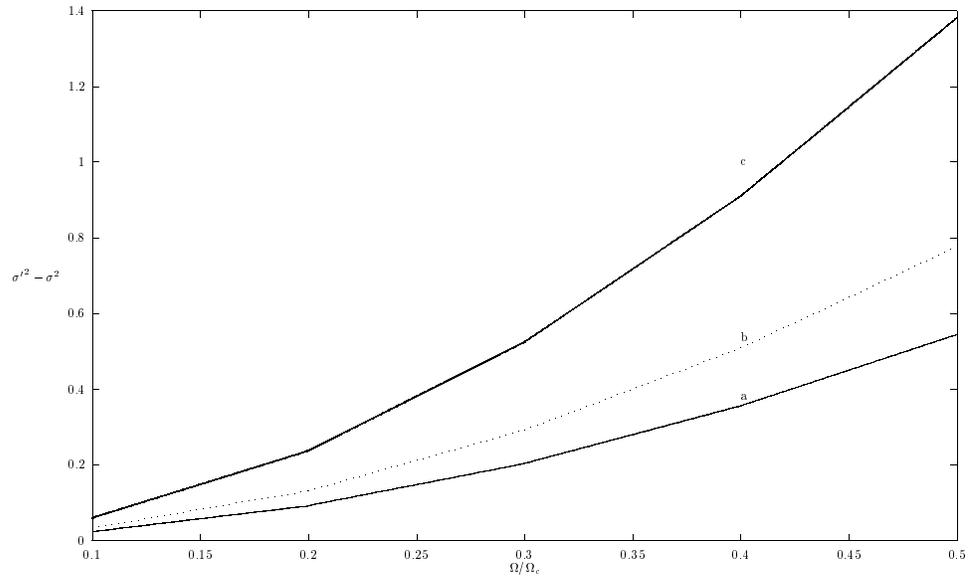}
\caption{(${\sigma^{\prime}}^2-{\sigma}^2$) in units of $9\times10^{10}sec.^{-2}$ vs $\Omega/\Omega_c$ for $n=0$. Curves a, b, c refer to maximum masses and corresponding magnetic fields: 2.17$M_{\odot}$, 0$Mev^2$; 3.0$M_{\odot}$, $1\times10^{5}Mev^{2}$ and 3.66$M_{\odot}$, $2\times10^{5}Mev^{2}$ respectively.}
\end{figure}

\begin{figure}[ht]
\vskip 15truecm
\includegraphics{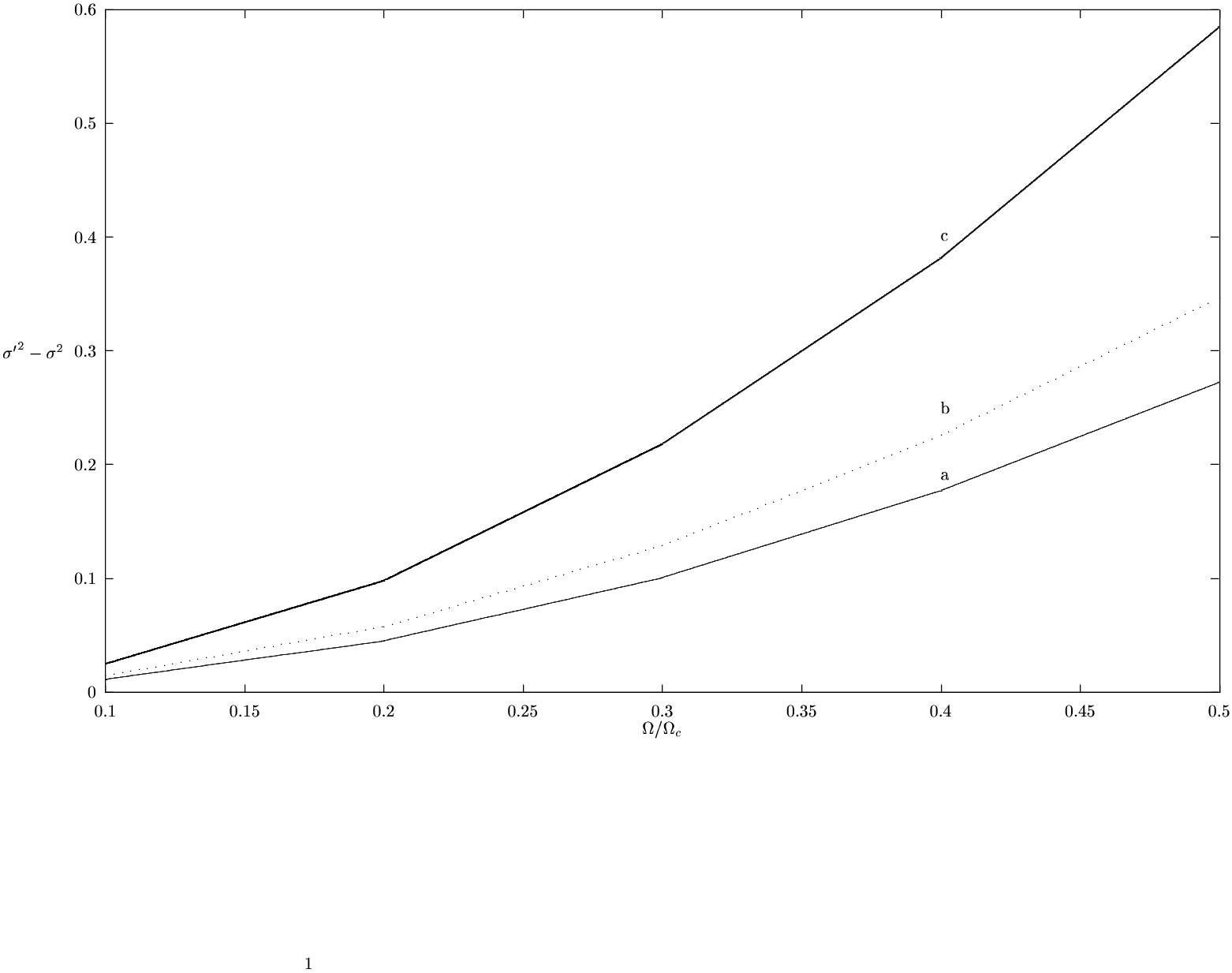}
\caption{Same as figure 9 for $n=1$.}
\end{figure}

\begin{figure}[ht]
\vskip 15truecm
\includegraphics{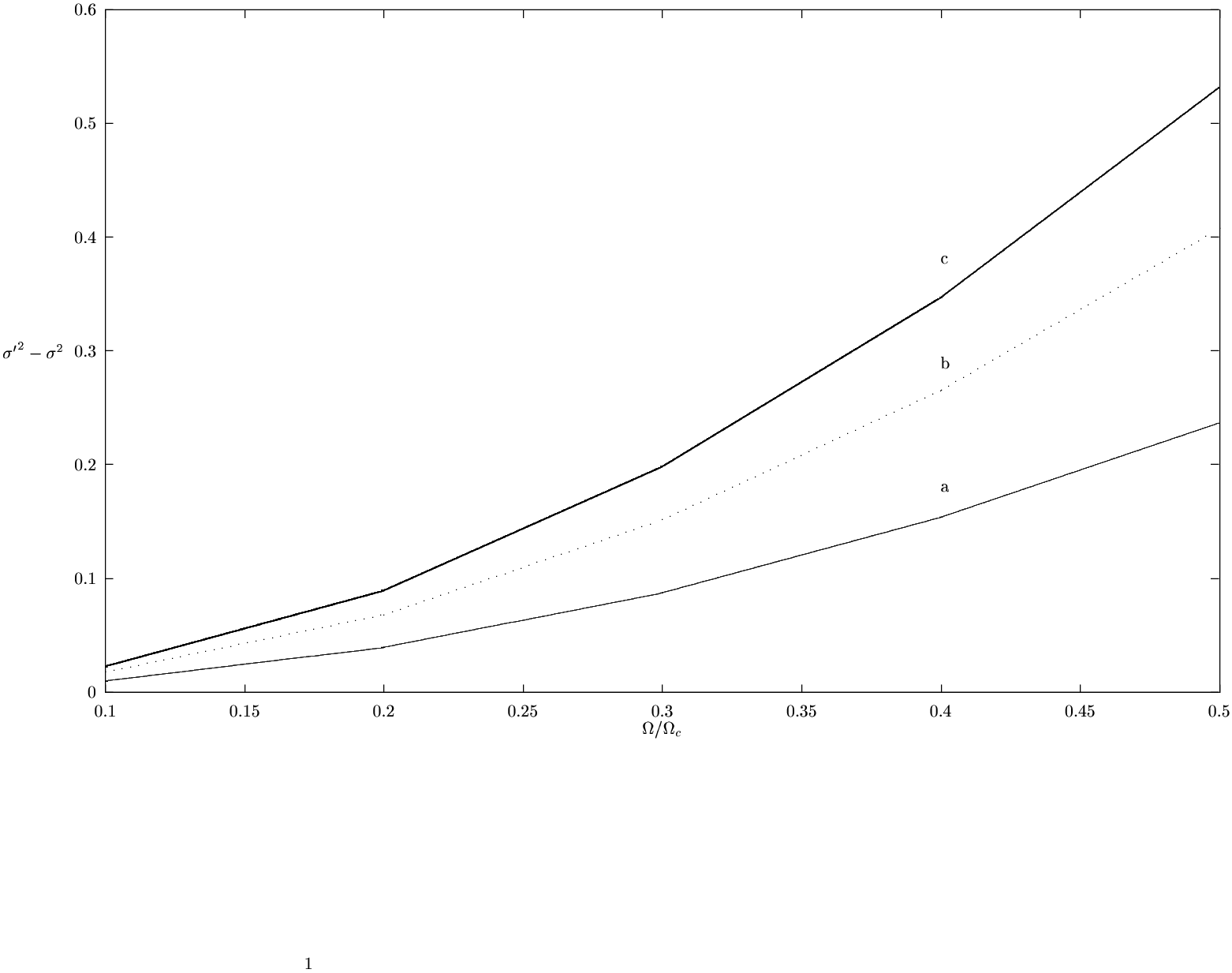}
\caption{Same as figure 9 for $n=2$.}
\end{figure}

\end{document}